\documentclass[prl,aps,showpacs,floatfix,twocolumn,superscriptaddress]{revtex4-2}
\usepackage{amsmath}
\usepackage{amsfonts}
\usepackage{amssymb}
\usepackage{revsymb}
\usepackage{graphicx}
\usepackage{hyperref}
\usepackage{dcolumn}
\usepackage{multirow}
\usepackage[table]{xcolor}
\usepackage{pdfpages}
\usepackage{physics}
\usepackage{mathtools}
\usepackage{tabularx}
\usepackage{booktabs}

\makeatletter \AtBeginDocument{\let\LS@rot\@undefined} \makeatother % <-- THE FIX
\graphicspath{{figs/}}

\newcommand{\eref}[1]{Eq.~(\ref{#1})}
\newcommand{\fref}[1]{Fig.~\ref{#1}}

\newcommand{\Ba}{$^{138}\mathrm{Ba}^+$}

\begin{document}

\title{Optimal state detection without likelihood estimation}
\author{M. D. K. Lee}%
\affiliation{Centre for Quantum Technologies, National University of Singapore, 3 Science Drive 2, Singapore 117543}%
\author{Z. Chai}
\affiliation{Department of Physics, National University of Singapore, 2 Science Drive 3, Singapore 117551}
\author{J. Z. F. Seng}
\affiliation{Department of Physics, National University of Singapore, 2 Science Drive 3, Singapore 117551}
\author{K. J. Arnold}%
\affiliation{Centre for Quantum Technologies, National University of Singapore, 3 Science Drive 2, Singapore 117543}%
\affiliation{Temasek Laboratories, National University of Singapore, 5A Engineering Drive 1, Singapore 117411}
\author{M.~D.~Barrett}
\affiliation{Centre for Quantum Technologies, National University of Singapore, 3 Science Drive 2, Singapore 117543}
\affiliation{Department of Physics, National University of Singapore, 2 Science Drive 3, Singapore 117551}

\begin{abstract}
We show that adaptive Bayesian estimation in fluorescence state detection reduces to a comparison of elapsed time with two arithmetic time sequences fixed by the bright and dark count rates and confidence parameters $\epsilon_d$ and $\epsilon_b$.  No probabilities are computed in real time and the resulting implementation, which we term sequential deadline detection, is provably optimal in mean detection time.  The analysis permits closed-form expressions for detection-time distributions, mean detection times, error rates of both channels accounting for miscalibrated rates, decay between the states during detection, and afterpulsing in single photon counting modules.  The predictions are validated by Monte Carlo simulation and demonstrated by fluorescence detection with \Ba{}.
\end{abstract}

\maketitle

High-fidelity discrimination of internal states is a basic requirement of trapped-ion quantum information processing and optical-clock metrology~\cite{Wineland1998,Ludlow2015}.  In the standard approach, state-dependent fluorescence is collected on a cycling transition so that an ion projected into one qubit state (``bright'') scatters many photons, while an ion in the other state (``dark'') scatters comparatively few~\cite{Nagourney1986,Sauter1986,Bergquist1986}.  The number of photons registered in a fixed detection window is then compared with a threshold to assign a state.  For finite collection efficiency and finite state lifetimes, however, the bright and dark photon-number distributions overlap, and a fixed threshold applied to a fixed window trades interrogation time against fidelity in a way that is far from optimal~\cite{Myerson2008,Langer2006}.

Adaptive and maximum-likelihood techniques improve on this by using the full temporal record of photon arrivals.  Rather than waiting a fixed time, one accumulates evidence and terminates as soon as a preset confidence level is reached~\cite{Myerson2008,Hume2007,Gehr2010}.  Bayesian detection is the natural formal setting: the posterior probability that the ion is bright is updated after each detection interval and the measurement is stopped when the posterior crosses a threshold~\cite{Myerson2008,Hume2007}.  The connection between adaptive readout and classical sequential analysis has been made explicit~\cite{DAnjou2016}, and photon arrival-time statistics have been exploited to speed up trapped-ion and atom readout~\cite{Crain2019,Geller2022}.  In all of these implementations, however, a likelihood function is computed and updated in real time, which imposes a nontrivial computational and timing overhead on field-programmable gate array (FPGA) implementations, particularly when many ions are read out in parallel~\cite{Burrell2010}.  The cost is real enough that demonstrations of trap-integrated detectors have either run the Bayesian analysis entirely in post-processing~\cite{Todaro2021} or noted that its speed advantage is offset by the computational resources it demands~\cite{McGuinness2022}.

In this paper we show that the Bayesian decision rule has an exceptionally simple form.  The decision rule collapses to a comparison of the elapsed time with two arithmetic sequences of deadlines, which are fixed by the bright and dark count rates and target confidence parameters $\epsilon_d$ and $\epsilon_b$.  A photon that misses one deadline registers a dark state, or one that meets the other registers a bright state, and the deadlines advance linearly if no declaration is made.  No probabilities need be stored or updated in real time.   Only comparisons of elapsed time against precomputed constants are required, which is ideally suited to FPGA logic.  The linear boundaries behind this rule are a classic result in sequential analysis~\cite{Dvoretzky1953,PeskirShiryaev2000}, and the deadlines appear in \cite{EpsteinSobel1955}, but fluorescence readout has followed the discrete-time Bayesian formulation in which the posterior is computed and stored~\cite{Myerson2008,Hume2007,DAnjou2016,Crain2019,Geller2022}; the deadline form has not been exploited, nor has the closed-form theory needed to design with it been given.

Bayesian detection updates the bright-state probability after each detection window by Bayes' rule with Poisson likelihoods for the photon count in that window, and terminates when a set confidence in either state is reached.  With rates $\lambda_b$ and $\lambda_d$ for the bright and dark states, successive updates maintain the functional form of the posterior, which depends only on the total number of photons $n$ and time $t$, independent of window size.  Starting from equal priors, the bright-state probability is
\begin{equation}
\label{Bayesian}
P(t;n)=\frac{1}{1+\left(\lambda_d/\lambda_b\right)^n e^{(\lambda_b-\lambda_d)t}}.
\end{equation}
The log-odds thus decrease linearly in time until a photon is detected, at which point they step discontinuously towards the bright state.  Detection terminates when $P$ takes a value $<\epsilon_d$, indicating a dark state, or $>(1-\epsilon_b)$, indicating a bright state; the confidence parameters need not be equal.

Because $P(t;n)$ is monotonic in either $t$ or $n$ with the other variable fixed, each threshold crossing corresponds to a boundary in the $(t,n)$ plane.  Solving $P(t;n)=\epsilon_d$ and $P(t;n)=1-\epsilon_b$ yields two arithmetic time sequences,
\begin{equation}
\label{tauseq}
\tau_n=\tau_{0,d}+n \tau,\qquad \bar{\tau}_n=-\tau_{0,b}+n \tau,
\end{equation}
with
\begin{equation}
\label{tau0}
\tau_{0,\alpha}=\frac{\ln(1-\epsilon_\alpha)-\ln\epsilon_\alpha}{\lambda_b-\lambda_d},\qquad \tau=\frac{\ln\lambda_b-\ln \lambda_d}{\lambda_b-\lambda_d}.
\end{equation}
The definition of $\tau$ establishes $(\lambda_d, \lambda_b)$ as a conjugate pair of the form $(u, u^*)$ satisfying $u e^{-u\tau} = u^* e^{-u^*\tau}$ and $u\tau < 1 < u^*\tau$: a relation that underpins the analysis.

The decision rule now takes a strikingly simple form.  A dark state is declared if the $n$th photon fails to make the deadline $\tau_{n-1}$, a bright state is declared if it beats the deadline $\bar{\tau}_n$, or the deadlines advance by $\tau$ if no declaration is made.  Consequently, probability distributions and likelihood estimates need not be evaluated at all.  The implementation need only maintain a photon counter $n$ and a timer, and compare the timer against the precomputed constants $\tau_{n-1}$ and $\bar\tau_n$.  As this is the sequential probability ratio test for two Poisson rates~\cite{Dvoretzky1953,PeskirShiryaev2000} with its linear boundaries solved for the deadline times \cite{EpsteinSobel1955}, we refer to the rule as sequential deadline detection.

The distribution of termination times for an atom that is genuinely dark or bright can be found using a \emph{single-boundary approximation} in which only the boundary that the state almost surely crosses is enforced.  The resulting distributions are exactly normalized but include the rare records that would have first crossed the opposing boundary.  The difference from the full two-boundary algorithm is of second order in the error rates.

\begin{figure}[t]
\centering
\includegraphics[width=\columnwidth]{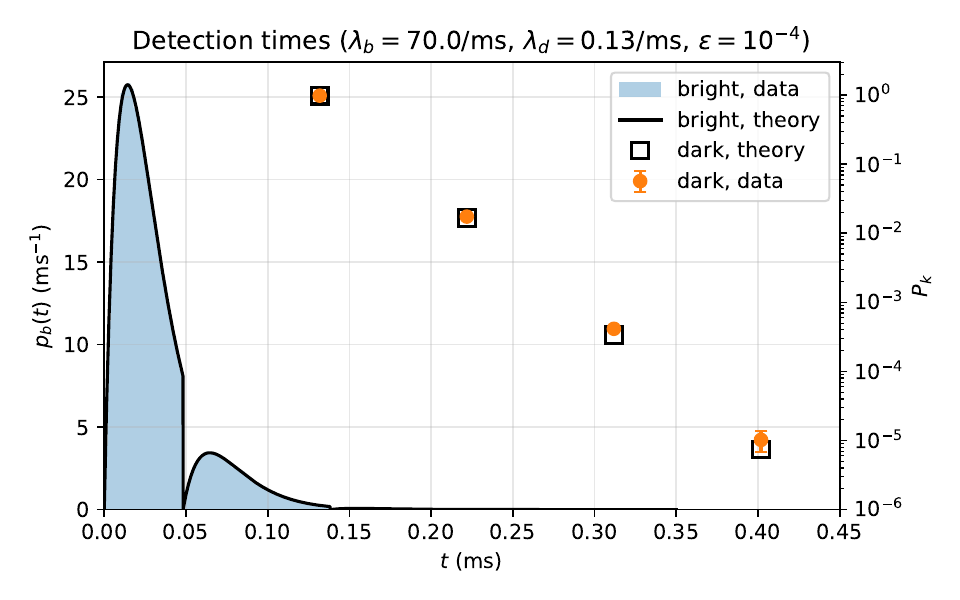}
\caption{\label{fig:tdist}
Measured detection-time distributions for \Ba{} at $\lambda_b=70$/ms, $\lambda_d=0.13$/ms and $\epsilon=10^{-4}$.  Bright state (left axis): histogram of $5\times10^5$ detections with \eref{brightP}; here $n_0=2$ and the step at $\bar\tau_2=48\,\mu$s separates two- and three-photon declarations.  Dark state (right axis): probabilities of termination at the deadlines $\tau_k$ from $8.8\times10^5$ detections (circles) with \eref{darkP} (open squares).}
\end{figure}

For a dark state, photons arrive as a Poisson process of rate $\lambda_d$, and termination occurs at one of the discrete times $\tau_k$, which has probability \cite{supp}
\begin{equation}
\label{darkP}
P_{k}=\frac{\tau_0 \lambda_d^{k} \tau_k^{k-1}e^{-\lambda_d \tau_{k}}}{k!},\qquad k\geq0.
\end{equation}
For a bright state, photons arrive as a Poisson process of rate $\lambda_b$ and termination occurs when the $n$th photon arrives before $\bar\tau_n=-\tau_{0,b}+n\tau$.  The smallest integer $n_0$ for which $\bar{\tau}_{n_0}>0$ is the minimum number of photons needed to detect a bright state.  The probability density of a detection time $t<\bar{\tau}_{n_0}$ is the Erlang distribution $E(t;n_0,\lambda_b)$, corresponding to the arrival of $n_0$ photons before $\bar{\tau}_{n_0}$.  For times $\bar{\tau}_{k}<t<\bar{\tau}_{k+1}$ with $k\ge n_0$, the probability density is \cite{Pyke1959,supp}
\begin{align}
\label{brightP}
p_{k+1}(t)
&= \lambda_b^{k+1} e^{-\lambda_b t}
\left(t-\bar{\tau}_k\right)\nonumber\\
&\qquad\times\sum_{m=0}^{n_0-1}
\frac{\bar{\tau}_m^m}{m!}
\frac{(t-\bar{\tau}_m)^{k-m-1}}{(k-m)!}.
\end{align}
The prefactor $(t-\bar{\tau}_k)$ makes the density vanish at the left edge of each interval, as it must: the $k$th photon must arrive after $\bar{\tau}_k$ yet before $t$.  Both distributions are shown in \fref{fig:tdist} for typical \Ba{} parameters, which lie in the regime where a dark state is usually declared with zero photons and a bright state with the minimum $n_0=2$.

\begin{figure}[t]
\centering
\includegraphics[width=0.95\linewidth]{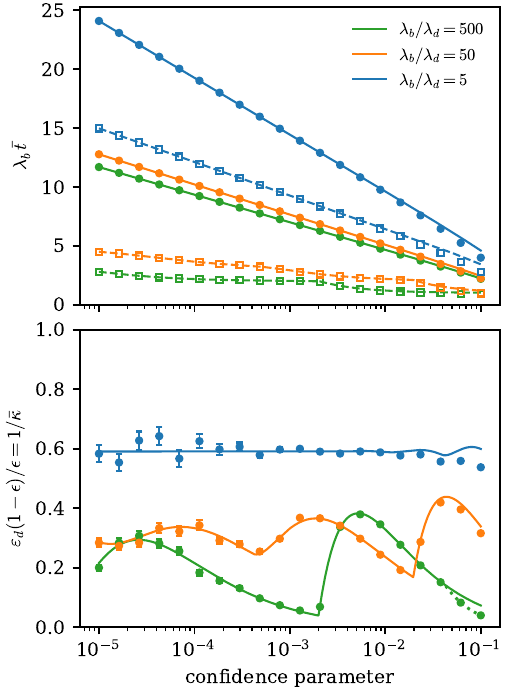}
\caption{\label{fig:mc}
Comparison with MC simulation of the full two-boundary algorithm for $\lambda_b/\lambda_d=5$, 50, and 500.  (Top) Mean detection times in photon units: circles and solid curves, dark ions; open squares and dashed curves, bright ions.  (Bottom) Dark-error suppression $\varepsilon_d(1-\epsilon_b)/\epsilon_b=1/\bar\kappa$: points with binomial error bars, MC; solid curves, \eref{edClosed}, whose arches follow the steps in $n_0$; dotted, $1-e^{-\lambda_d\tau_{0,b}}$ where $\tau_{0,b}<\tau/2$ and only a single photon race remains.  Agreement is within statistics for $\epsilon\lesssim10^{-2}$; the resolved deviations at larger $\epsilon$ are the second-order corrections neglected in the single-boundary approximation.}
\end{figure}

Mean detection times follow directly from the distributions.  For a dark state, $\bar{t}_d=\tau_{0,d}/(1-\lambda_d\tau)$, and, for a bright state, Wald's identity~\cite{Wald1944,Grimmett2001} gives the exact relation $\bar{t}_b=(\tau_{0,b}+\bar{o})/(\lambda_b\tau-1)$, where $\bar{o}\in(0,\tau)$ is the mean time by which the terminating photon precedes its deadline.  A closed form for $\bar{o}$ is given in \cite{supp}.  In either case the mean detection time grows only logarithmically with $1/\epsilon_\alpha$ through $\tau_{0,\alpha}$ as illustrated in \fref{fig:mc} over a wide range of contrasts and confidence parameters, and verified by Monte Carlo (MC) simulation of the full two-boundary algorithm.

The detection times are optimal.  The scheme is the sequential probability ratio test for discriminating two Poisson rates, and by the Wald--Wolfowitz theorem~\cite{WaldWolfowitz1948}, extended to continuous-time processes including the Poisson case~\cite{Dvoretzky1953,PeskirShiryaev2000}, no procedure, adaptive or otherwise, whose error probabilities do not exceed those achieved by the thresholds has a smaller mean detection time under either hypothesis.  The optimality is relative to the error pair actually achieved, which can also be determined analytically.

The distributions \eref{darkP} and \eref{brightP} are normalized but ignore the possibility of a false declaration: (i) A {\it bright error} with probability $\varepsilon_b$ in which a bright ion is falsely declared dark, or (ii) a {\it dark error} with probability $\varepsilon_d$ in which a dark ion is falsely declared bright.  Either error can be determined using the single-boundary approximation.

The thresholds enter \eref{darkP} and \eref{brightP} only through the fixed constants $\tau_{0,\alpha}$ and $\tau$, so both distributions hold for an arbitrary arrival rate $\lambda$.  For bright rates with $\lambda \tau> 1$ and dark rates with $\lambda \tau<1$, the probability distributions are normalized as the posterior must eventually cross the one boundary imposed.  When the condition is violated, crossing the boundary is no longer guaranteed but the total probability given by the distributions still represents the probability of doing so.  Consequently, interchanging the rates associated with each distribution provides estimates of the errors, which holds even if the true rates $\lambda_d'$ and $\lambda_b'$ are not the design rates $\lambda_d$ and $\lambda_b$, so long as $\lambda'_d\tau <1<\lambda'_b\tau$.  Additionally, since a record contributing to the difference from the true error must cross both boundaries, the evaluations are accurate to second order in the errors.

A bright error requires the record of a rate-$\lambda'_b$ process to miss a dark deadline.  Summing the dark-declaration probabilities \eref{darkP} at rate $\lambda'_b$ and using conjugacy gives
\begin{equation}
\label{ebBound}
\varepsilon'_b \simeq \sum_{n} P_n\Big|_{\lambda_d\to\lambda'_b}
=e^{-(\lambda'_b-\lambda_b^{\prime*})\tau_{0,d}}\xrightarrow{\lambda'_b=\lambda_b}\frac{\epsilon_d}{1-\epsilon_d},
\end{equation}
independent of the contrast.  Similarly, for the dark error at rate $\lambda'_d$,
\begin{equation}
\label{kdid}
\bar\kappa'\,\varepsilon'_d=e^{-(\lambda_d^{\prime*}-\lambda'_d)\tau_{0,b}}\xrightarrow{\lambda'_d=\lambda_d}\frac{\epsilon_b}{1-\epsilon_b},
\end{equation}
where $\bar\kappa'=\mathbb{E}_{d'}[\kappa']\ge1$ is the mean of $\kappa'=e^{(\lambda_d^{\prime*}-\lambda'_d)(\bar\tau_n-t)}$.  So $\bar{\kappa}'$ is the mean overshoot of the true likelihood-ratio process, which coincides with that of the model posterior only at the design point $\lambda'_d=\lambda_d$.  At the design rates, the dark error is thus suppressed below $\epsilon_b/(1-\epsilon_b)$ by the mean overshoot.  In general, $\bar{\kappa}'=e^{(\lambda_d^{\prime*}-\lambda'_d)\bar{o}_\mathrm{v}}$ for some $\bar{o}_\mathrm{v}(\lambda'_d)\in (0,\tau)$, but $\bar{\kappa}'$ and hence $\bar{o}_\mathrm{v}$ can be determined in closed form by direct integration.  Alternatively, the crossing condition maps onto the first passage of a random walk whose supremum is distributed as the stationary waiting time of an M/D/1 queue, and Erlang's waiting-time formula~\cite{Kleinrock1975} gives
\begin{equation}
\label{edClosed}
\varepsilon'_d\simeq1-(1-\lambda'_d\tau)\sum_{m=0}^{n_0-1}\frac{(\lambda'_d\bar\tau_m)^m}{m!}\,e^{-\lambda'_d\bar\tau_m}.
\end{equation}
We then have
\begin{equation}
\bar\kappa'=\frac{e^{-(\lambda_d^{\prime*}-\lambda'_d)\tau_{0,b}}}{\varepsilon'_d},\quad\mbox{and}\quad \bar{o}_\mathrm{v}=\frac{\ln\bar\kappa'}{\lambda_d^{\prime*}-\lambda'_d}.
\end{equation}
The relation is verified by MC simulation of the full two-boundary algorithm in \fref{fig:mc}.

In practice the true rates $\lambda_b'$ and $\lambda_d'$ will only differ marginally from the design values.  First order scaling of the errors can then be determined by expanding the conjugates about the design values, which gives sensitivities,
\begin{equation}
\label{mismatchlin}
\frac{\varepsilon_b'}{\varepsilon_b}\approx\left(\frac{\lambda_b'}{\lambda_b}\right)^{-(\lambda_b-\lambda_d)\bar t_d},\quad
\frac{\varepsilon_d'}{\varepsilon_d}\approx\left(\frac{\lambda_d'}{\lambda_d}\right)^{(\lambda_b-\lambda_d)\tilde t_b},
\end{equation}
where $\tilde t_b=(\tau_{0,b}+\bar o_\mathrm{s})/(\lambda_b\tau-1)$ is the mean bright detection time with $\bar o$ replaced by a closely related effective overshoot $\bar o_\mathrm{s}$, given in closed form in \cite{supp}.  Each error thus scales with the log-odds accumulated over the mean detection time of the \emph{opposite} state.  Typically $\tilde t_b\sim\bar{t}_b$, so $\varepsilon_d$ is the more robust of the two by the factor $\sim\bar t_b/\bar t_d$.

%\begin{figure}[t]
%\centering
%\includegraphics[width=\columnwidth]{fig_mismatch.pdf}
%\caption{\label{fig:mismatch}
%Bright and dark error rates as the true bright and dark detection rates deviate from the model parameters, shown as a function of the confidence %parameter $\epsilon$.  [Figure to be supplied.]}
%\end{figure}

Errors are also limited by the states themselves through decay during detection.  To leading order, this may be treated as one-way decays of each channel, which add independently.  If the state is undecided at the point of decay, the declaration follows the new state unless the posterior is already near the original threshold.  The additional error is the probability of decay within the mean detection time, offset by a fixed time that accounts for the near-boundary exception, giving \cite{supp}
\begin{align}
\Delta\varepsilon_d&\simeq\gamma\left[\bar{t}_d-\lambda_b/(\lambda_b-\lambda_d)^2\right],\\
\label{decayerr}
\Delta\varepsilon_b&\simeq\gamma\left[\bar{t}_b-\lambda_d/(\lambda_b-\lambda_d)^2\right],
\end{align}
for the dark and bright states, respectively, where $\gamma$ is the decay rate for its respective channel.  The decay contribution is fully determined by the rates, and because the time-optimal test minimizes $\bar{t}_\alpha$, it minimizes this channel as well.  In \Ba{}, the dark state is limited by quadrupole decay of the shelved $D_{5/2}$ state.  For the rates of \fref{fig:tdist} and the $\sim30$\,s lifetime, \eref{decayerr} gives $\Delta\varepsilon_d\approx5\times10^{-6}$ at $\epsilon=10^{-5}$, growing only logarithmically as $\epsilon$ is reduced, so the total dark error passes through a soft minimum near the $10^{-5}$ level.  Before this limit is reached, however, the dark error is dominated by afterpulsing, a common limitation of single photon counting modules (SPCMs).

Afterpulsing duplicates a photon count and increases the dark error.  When the dark error is dominated by the Erlang tail, it is given by $\varepsilon_d\approx(\lambda_d \bar{\tau}_{n_0})^{n_0}/n_0!$. To first order in the afterpulse probability $p$, the dark error correction is then approximated by
\begin{equation}
\label{APCorr}
\Delta \varepsilon_d=(n_0-1)p\frac{(\lambda_d \bar{\tau}_{n_0})^{n_0-1}}{(n_0-1)!}.
\end{equation} 
The correction is the probability that $n_0-1$ real photons land within a time $\bar\tau_{n_0}$ and one of them is trailed by an afterpulse still inside the window; afterpulse delays are instantaneous compared to typical values of $\bar\tau_{n_0}$.  A more general form is given in \cite{supp} but \eref{APCorr} typically applies when afterpulsing is a significant limitation.  

\begin{figure}[t]
\centering
\includegraphics[width=\columnwidth]{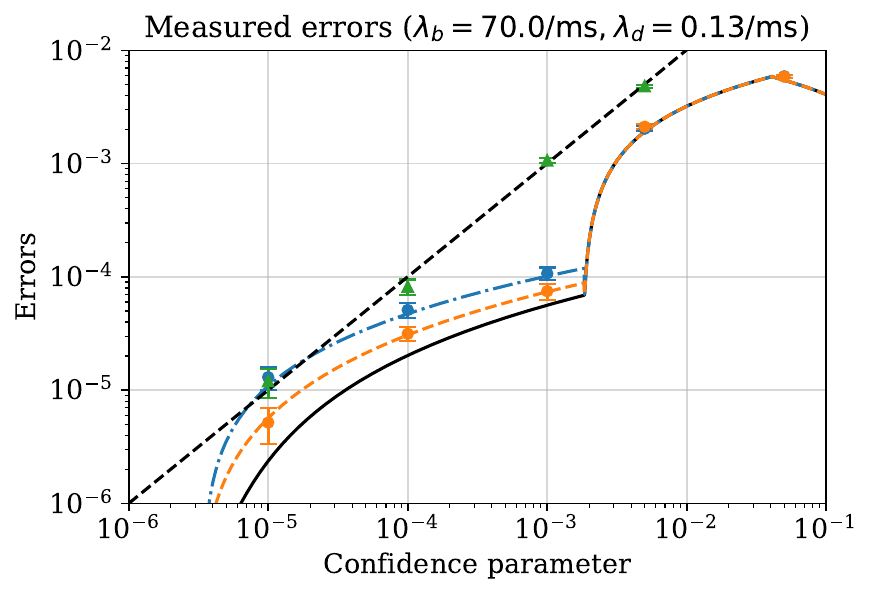}
\caption{\label{fig:apdark}
Measured error rates versus confidence parameter $\epsilon=\epsilon_d=\epsilon_b$ for \Ba{} at $\lambda_b=70$/ms and $\lambda_d=0.13$/ms.  Dark error with the native SPCM dead time ($\Delta=0$, blue circles) and with counts blanked for $\Delta=1\,\mu$s after each detection (orange squares); curves are \eref{edClosed} alone (solid; the single-race form $1-e^{-\lambda_d\tau_{0,d}}$ for $\epsilon>0.04$, where $\tau_{0,d}<\tau/2$) and with the afterpulse correction \eref{APCorr} at the independently measured $p=0.43\%$ (dash-dotted) and $0.17\%$ (dashed).  Bright error (green triangles) with \eref{ebBound}, $\varepsilon_b=\epsilon_d/(1-\epsilon_d)$ (black dashed).  Decay events are excluded as described in the text.}
\end{figure}

Figure~\ref{fig:apdark} shows the measured error probabilities at the rates of \fref{fig:tdist}.  The bright error follows $\epsilon_d/(1-\epsilon_d)$ over the full range, as \eref{ebBound} requires, independent of contrast and unaffected by afterpulsing, which can only advance a bright declaration.  The dark error is compared with \eref{edClosed} and \eref{APCorr}.  To isolate the afterpulse channel, the ion is optically pumped to $D_{5/2}$ and detected repeatedly until the state decays.  Decays must be separated from errors, which is done by confirmation: each bright declaration is followed by a second detection.  As two consecutive dark errors are improbable, a repeated bright declaration flags a decay and ends the run, whereas a subsequent dark declaration flags a genuine error.  The afterpulse probability is obtained independently from the photon inter-arrival times.  An exponential fit to delays well beyond a microsecond captures the Poisson statistics of the fluorescence, and the excess of the full distribution over this fit gives $p=0.43\%$.  The intrinsic SPCM dead time of $\sim15$\,ns plays no role at these count rates, but extending it to 1\,$\mu$s, by disregarding any count within that interval, suppresses the afterpulses giving the second data set with $p=0.17\%$.  Neither the closed form nor the correction contains a free parameter.  Above the step at $\epsilon\simeq2\times10^{-3}$, where a single photon suffices for a bright declaration, afterpulsing has negligible effect and the data confirm the uncorrected error.  Below it, the correction is essential, exceeding a factor of two at $\epsilon=10^{-4}$, and blanking reduces the excess in proportion to $p$, as expected for a first-order effect.  

In summary, we have shown that Bayesian state detection reduces to sequential deadline detection, a simple adaptive rule based entirely on photon arrival times.  The decision boundaries collapse onto two arithmetic sequences $\tau_n=\tau_{0,d}+n\tau$ and $\bar\tau_n=-\tau_{0,b}+n\tau$.  The state is declared dark if the $n$th photon misses the $\tau_{n-1}$ deadline, or bright if it meets $\bar\tau_n$.  The hardware requirement is a counter, a timer, and two stored constants, which is no more than threshold detection already uses. The scheme also delivers photon arrival times as a by-product; where photon-correlation techniques apply \cite{berkeland1998minimization,keller2015precise}, routine detections correlated with the trap drive phase then accumulate a record of micromotion at no additional cost.  Any fluorescence-based readout, including the parallel readout of ion and atom arrays, can therefore adopt the rule without additional complexity, and the closed-form detection times and error probabilities, including the effects of rate mismatch, decay, and afterpulsing, provide the design tools to quantify performance.  For readout limited by Poissonian scattering and state decay, no procedure with the same error rates can decide faster on average, and we see no setting in which the scheme would not be the preferred choice.
\begin{acknowledgments}
This project is supported by the National Research Foundation, Singapore through the National Quantum Office, hosted in A*STAR, under its Quantum Engineering Programme 3.0 Funding Initiative (W25Q3D0007) and under its Centre for Quantum Technologies Funding Initiative (S24Q2d0009).
\end{acknowledgments}
\bibliography{detection}
\clearpage % Finish the main text cleanly
% This loops through your PDF pages without breaking REVTeX's page dimensions
\onecolumngrid 



\section*{Supplementary material}

# Supplemental Material: Optimal state detection without likelihood estimation

M. D. K. Lee,[1] Z. Chai,[1] J. Z. F. Seng,[1] K. J. Arnold,[1,2] and M. D. Barrett[1,3]

[1]Centre for Quantum Technologies, National University of Singapore, 3 Science Drive 2, Singapore 117543
[2]Temasek Laboratories, National University of Singapore, 5A Engineering Drive 1, Singapore 117411
[3]Department of Physics, National University of Singapore, 2 Science Drive 3, Singapore 117551

This Supplemental Material gives the mathematical derivations of results given in the main text. Derivations are either relatively straightforward or have appeared in some form within the literature, but have been compiled here for the convenience of the interested reader.

## I. DARK STATE ANALYSIS

### A. Probability Distribution

For a dark ion, photons arrive as a Poisson process of rate $\lambda_d$, and termination occurs at one of the discrete times $\tau_n$. The algorithm terminates at $\tau_n$ if exactly $n$ photons have been registered, with the $(k+1)$th photon arriving before $\tau_k$ for every $k<n$ so that no earlier dark declaration occurred, and with no photon in the final interval of length $\tau_n$ following the $n$th photon. Denoting the probability of a detection time $\tau_k$ by $P_k$, we have

$$P_k = \lambda_d^k e^{-\lambda_d \tau_k} \int_0^{\tau_0}\int_{t_1}^{\tau_1}\cdots\int_{t_{k-1}}^{\tau_{k-1}} dt_k\, dt_{k-1}\cdots dt_1 \qquad \text{(S1)}$$

for $k>0$ and $P_0 = e^{-\lambda_d \tau_0}$.

The dark distribution Eq. (4) follows from

$$I_k \equiv \int_0^{\tau_0}\int_{t_1}^{\tau_1}\cdots\int_{t_{k-1}}^{\tau_{k-1}} dt_k \cdots dt_1 = \frac{\tau_0\, \tau_k^{k-1}}{k!}, \qquad \text{(S2)}$$

valid for the arithmetic sequence $\tau_j = \tau_0 + j\tau$, applied to the constraint that the $(j+1)$th photon arrive before $\tau_j$ for every $j<k$. The cases $k=1,2$ are immediate: $I_1 = \tau_0$ and

$$I_2 = \int_0^{\tau_0} (\tau_1 - t_1)\, dt_1 = \frac{\tau_0(\tau_0+2\tau)}{2} = \frac{\tau_0\tau_2}{2}. \qquad \text{(S3)}$$

For the general case, carrying out the innermost integral and proceeding inductively generates a recursion whose solution is $\tau_0\tau_k^{k-1}/k!$. Equivalently, Eq. (S2) is the volume of the Pitman–Stanley polytope [1]. This gives Eq. 4 of the main text for any $k\geq 0$.

### B. Normalization and bright error rates

Normalization and error rates follow from the generalized tree series

$$\sum_{n\geq 0} \frac{a(a+nc)^{n-1}y^n}{n!} = e^{aT}, \qquad \text{(S4)}$$

where $T=T(y)$ satisfies $T = y\, e^{cT}$, i.e. $cT = -\mathrm{W}(-cy)$ and W is the Lambert W function [2]. Identifying $a=\tau_0$, $c=\tau$, and $y=\lambda e^{-\lambda\tau}$ gives

$$\sum_{k\geq 0} P_k = \exp\Big[\lambda\tau_0 - \frac{\tau_0}{\tau}\,\mathrm{W}\big(-\lambda\tau\, e^{-\lambda\tau}\big)\Big]. \qquad \text{(S5)}$$

Since $ue^{-u}\leq e^{-1}$, the argument $-\lambda\tau e^{-\lambda\tau}$ always lies in $[-e^{-1},0)$ and the series converges unconditionally, summing to the principal branch: $-\mathrm{W}(-\lambda\tau e^{-\lambda\tau})$ equals $\lambda\tau$ itself for $\lambda\tau\leq 1$, and the smaller conjugate solution of $ue^{-u} = \lambda\tau e^{-\lambda\tau}$ for $\lambda\tau>1$. Normalization then follows for any $\lambda_d'\tau<1$ and the bright error rates for any $\lambda_b'\tau>1$.

### C. Mean detection time

The companion identity [3]

$$\sum_{n\geq 0}\frac{(a+nc)^n y^n}{n!} = \frac{e^{aT}}{1-cT} \qquad \text{(S6)}$$

yields the mean detection time

$$\bar{t}_d = \frac{\tau_0}{1-\lambda\tau} \qquad \text{(S7)}$$

valid for any $\lambda\tau<1$ and not dependent on $\lambda$ being the design rate $\lambda_d$.

## II. BRIGHT STATE ANALYSIS

### A. Probability distribution

For a bright ion, photons arrive as a Poisson process of rate $\lambda_b$, and the $n$th photon triggers a bright declaration when it arrives before $\bar{\tau}_n = -\tau_0 + n\tau$. For convenience $\tau_0 = \tau_{0,b}$ for this section. The smallest integer $n_0$ for which $\bar{\tau}_{n_0}>0$ is the minimum number of photons needed to detect a bright state. To obtain a detection time $\bar{\tau}_{n-1}<t<\bar{\tau}_n$, there must be $n$ photons, with the $k$th photon having an arrival time $t_k>\bar{\tau}_k$ for every $k<n$ so that no earlier bright declaration occurred. The probability density of a detection time $t<\bar{\tau}_{n_0}$ is the Erlang distribution $E(t;n_0,\lambda_b)$, corresponding to the arrival of $n_0$ photons before $\bar{\tau}_{n_0}$. For times $\bar{\tau}_k<t<\bar{\tau}_{k+1}$ with $k\geq n_0$, the density is

$$p_{k+1}(t) = \lambda_b^{k+1} e^{-\lambda_b t} V_k(t) \qquad \text{(S8)}$$

where

$$V_k(t) = \frac{1}{(n_0-1)!} \times \int_{\bar\tau_k}^{t}\int_{\bar\tau_{k-1}}^{t_k}\cdots\int_{\bar\tau_{n_0}}^{t_{n_0+1}} t_{n_0}^{n_0-1}\,dt_{n_0}\cdots dt_k, \quad \text{(S9)}$$

is the volume of ordered arrival times $0 < t_1 < \cdots < t_k < t$ subject to $t_j \geq \bar\tau_j$ for all $j$. The constraints with $j < n_0$ are vacuous since $\bar\tau_j < 0 < t_j$. The lower limits of integration for these variables are zero and integrating them out produces the factor $t_{n_0}^{n_0-1}/(n_0-1)!$. Differentiation with respect to the upper limit gives $V_k'(t) = V_{k-1}(t)$, subject to the boundary data

$$V_k(t) = \frac{t^k}{k!} \quad (k < n_0), \qquad V_k(\bar\tau_k) = 0 \quad (k \geq n_0). \quad \text{(S10)}$$

The derivative structure is solved by an expansion on the Abel basis with $k$-independent coefficients,

$$V_k(t) = \sum_{j=0}^{k} b_j\,\frac{(t-\bar\tau_j)^{k-j}}{(k-j)!}, \quad \text{(S11)}$$

and Eq. (S10) determines the $b_j$ recursively. For $k < n_0$ the Abel binomial identity [3]

$$\sum_{k=0}^{n}\binom{n}{k} x(x+kz)^{k-1}(y-kz)^{n-k} = (x+y)^n, \quad \text{(S12)}$$

with $x = -\tau_0$, $z = \tau$, and $y = t + \tau_0$, gives $b_j = -\tau_0\bar\tau_j^{\,j-1}/j!$ for $0 \leq j \leq n_0 - 1$. For $k \geq n_0$, evaluating Eq. (S11) at $t = \bar\tau_k$ and using $\bar\tau_k - \bar\tau_j = (k-j)\tau$ turns the second condition in Eq. (S10) into the statement that the series $B(z)F(z)$, with $B(z) = \sum_j b_j z^j$ and

$$F(z) = \sum_{m\geq 0}\frac{(m\tau)^m}{m!}\,z^m = \frac{1}{1-T(\tau z)}, \quad \text{(S13)}$$

terminates at order $n_0-1$. Hence $B(z) = Q(z)[1-T(\tau z)]$ for a polynomial $Q$ of degree $n_0 - 1$. Noting

$$[z^m]\big(B(z)F(z)\big) = \sum_{j=0}^{m} b_j\,\frac{\big((m-j)\tau\big)^{m-j}}{(m-j)!}, \quad \text{(S14)}$$

and $m \leq n_0 - 1$, so that $b_j = -\tau_0\bar\tau_j^{\,j-1}/j!$, we have,

$$\sum_{j=0}^{m}\frac{x(x+j\tau)^{j-1}}{j!}\,\frac{(m\tau-j\tau)^{m-j}}{(m-j)!} = \frac{1}{m!}\sum_{j=0}^{m}\binom{m}{j} x(x+j\tau)^{j-1}(y-j\tau)^{m-j}, \quad \text{(S15)}$$

with $x = -\tau_0$ and $y = m\tau$. Abel's identity Eq. (S12) then gives

$$Q(z) = \sum_{m=0}^{n_0-1}\frac{\bar\tau_m^{\,m}}{m!}\,z^m. \quad \text{(S16)}$$

The generating function of Eq. (S11) is

$$\sum_{k\geq 0} V_k(t)\,z^k = e^{(t+\tau_0)z}\,B(ze^{-\tau z}) = e^{(t+\tau_0)z}\,(1-\tau z)\sum_{m=0}^{n_0-1}\frac{\bar\tau_m^{\,m}}{m!}\,z^m e^{-m\tau z}, \quad \text{(S17)}$$

where the substitution $z \to ze^{-\tau z}$ arises from $e^{-\bar\tau_j z} = e^{\tau_0 z}(e^{-\tau z})^j$ and the tree-function identity $T(we^{-w}) = w$ gives $B(ze^{-\tau z}) = (1-\tau z)Q(ze^{-\tau z})$. Reading off the coefficient of $z^k$ and using $\bar\tau_m + (k-m)\tau = \bar\tau_k$ gives

$$V_k(t) = (t-\bar\tau_k)\sum_{m=0}^{n_0-1}\frac{\bar\tau_m^{\,m}}{m!}\,\frac{(t-\bar\tau_m)^{k-m-1}}{(k-m)!}, \quad \text{(S18)}$$

and Eq. (5) then follows. The result is contained in Lemma 1 of [4].

### B. Normalization and dark error rates

For an arbitrary rate $\lambda$, let $u_k(t) = \lambda^k e^{-\lambda t}V_k(t)$, so that the interval densities are $p_{k+1} = \lambda u_k$ and $V_k' = V_{k-1}$ gives $u_k' = \lambda(u_{k-1}-u_k)$. For $\bar\tau_K < t < \bar\tau_{K+1}$, the survival sum

$$\Sigma(t) = \sum_{k=0}^{K} u_k(t) \quad \text{(S19)}$$

telescopes to $\Sigma'(t) = -\lambda u_K(t)$ and is continuous across the interval boundaries, since each entering term vanishes there by Eq. (S10). At $t = 0$ only indices $k \leq K = n_0 - 1$ appear, where $V_k(t) = t^k/k!$, so $\Sigma(0) = V_0(0) = 1$ and

$$\int_0^t p(t')\,dt' = 1 - \Sigma(t) \quad \text{(S20)}$$

for all $t$, where $p(t)$ denotes the piecewise density including its Erlang segment, so the total crossing probability is $1 - \Sigma(t)$ for $t \to \infty$.

The limit follows from Eq. (S11). Interchanging the sums gives,

$$\Sigma(t) = e^{\lambda\tau_0}\sum_{j=0}^{K} b_j\big(\lambda e^{-\lambda\tau}\big)^j\,\Pi_{K-j}\big(\lambda(t-\bar\tau_j)\big), \quad \text{(S21)}$$

where

$$\Pi_n(\mu) = e^{-\mu}\sum_{r\leq n}\frac{\mu^r}{r!}$$

is the probability that a Poisson variable of mean $\mu$ does not exceed $n$. With $x = \lambda e^{-\lambda\tau}$ and fixed $J$

$$\Sigma(t) = e^{\lambda\tau_0}\Big[\sum_{j=0}^{J} b_j x^j\,\Pi_{K-j}\big(\lambda(t-\bar\tau_j)\big) + \sum_{j=J+1}^{K} b_j x^j\,\Pi_{K-j}\big(\lambda(t-\bar\tau_j)\big)\Big]. \quad \text{(S22)}$$

Since $0 \leq \Pi_n \leq 1$ and $B(x)$ is absolutely convergent, the second sum can be made arbitrarily small for sufficiently large $J$. For $j \leq J$, $\lambda(t-\bar{\tau}_j)/(K-j) \to \lambda\tau$, so $\Pi_{K-j}\big(\lambda(t-\bar{\tau}_j)\big)$ tends to one for $\lambda\tau < 1$ and to zero for $\lambda\tau > 1$. Hence the survival sum vanishes for $\lambda\tau > 1$, which establishes normalization, and, for $\lambda\tau < 1$, $B(\lambda e^{-\lambda\tau}) = (1-\tau\lambda)Q(\lambda e^{-\lambda\tau})$ gives

$$\lim_{t\to\infty} \Sigma(t) = (1-\lambda\tau)\sum_{m=0}^{n_0-1} \frac{(\lambda\bar{\tau}_m)^m}{m!}\, e^{-\lambda\bar{\tau}_m}, \qquad \text{(S23)}$$

which is Erlang's waiting-time distribution: with $\lambda \to \lambda'_d$ the crossing probability $1-\Sigma(\infty)$ gives Eq. (8). The survival probability is $\sigma(x,T)$ given in [4] where the same result is established.

### C. Overshoot distribution and mean detection time

The mean detection time for a bright state is more cumbersome than the dark counterpart. Under the bright hypothesis, the analogous drift toward the bright threshold has mean rate $(\lambda_b-\lambda_d)(\lambda_b\tau-1)$, but the boundary is crossed by a jump. With $N$ denoting the photon number at which detection terminates, so that the $N$th photon arrives at time $t < \bar{\tau}_N$, and $o = \bar{\tau}_N - t \in (0,\tau)$ the time by which it precedes its threshold, Wald's identity $\bar{t}_b = \mathbb{E}[N]/\lambda_b$ gives the exact relation

$$\bar{t}_b = \frac{\tau_0 + \bar{o}}{\lambda_b\tau - 1}. \qquad \text{(S24)}$$

The density, $f(o)$, of the overshoot $o = \bar{\tau}_{k+1} - t$ at a bright declaration is given by

$$f(o) = \sum_{k\geq n_0-1} p_{k+1}(\bar{\tau}_{k+1} - t)$$

requiring the diagonal sum $\sum_k x^{k+1}V_k(\bar{\tau}_{k+1}-o)$ with $x = \lambda_b e^{-\lambda_b\tau}$. Substituting $t = \bar{\tau}_{k+1} - o$ in Eq. (S17) gives

$$V_k(\bar{\tau}_{k+1}-o) = [z^k]\big(e^{k\tau z}g(z)\big),$$

where $[z^k]$ denotes the coefficient of $z^k$ and

$$g(z) = e^{(\tau-o)z}(1-\tau z)Q(ze^{-\tau z}).$$

Lagrange inversion of $x = \zeta e^{-\tau\zeta}$ about $x=0$ gives

$$\sum_{k\geq 0} x^k [z^k]\big(e^{k\tau z}g(z)\big) = g(\zeta)/(1-\tau\zeta),$$

where $\zeta = \lambda_d$ is the conjugate of $\lambda_b$. Multiplying by $e^{\lambda_b(\tau_0+o)}$, subtracting the unphysical $k < n_0-1$ terms, and restricting the $k = n_0-1$ term to $o < \bar{\tau}_{n_0}$ yields,

$$f(o) = \lambda_d\, e^{\lambda_b\tau_0}\, Q\big(\lambda_b e^{-\lambda_b\tau}\big)\, e^{(\lambda_b-\lambda_d)o} - e^{\lambda_b(\tau_0+o)} {\sum_{j=1}^{n_0}}' \left(\lambda_b e^{-\lambda_b\tau}\right)^j \frac{(\bar{\tau}_j - o)^{j-1}}{(j-1)!}, \qquad \text{(S25)}$$

for $o \in (0,\tau)$, where the prime indicates that the $j = n_0$ term is present only for $o > \bar{\tau}_{n_0}$; up to the boundary terms the density is a pure exponential in the log-odds overshoot. Defining $A = \lambda_d e^{\lambda_b\tau_0}Q(\lambda_b e^{-\lambda_b\tau})$ and $\beta = \lambda_b - \lambda_d$, term-by-term integration yields,

$$\bar{o} = \frac{A}{\beta^2}\left[1 + e^{\beta\tau}(\beta\tau - 1)\right] - \sum_{j=1}^{n_0} \bar{o}_j, \qquad \text{(S26)}$$

where

$$\bar{o}_j = \frac{\bar{\tau}_j}{(j-1)!}\Big[\gamma(j,\lambda_b\bar{\tau}_j) - \gamma(j,\lambda_b\bar{\tau}_{j-1})\Big] - \frac{1}{\lambda_b(j-1)!}\Big[\gamma(j+1,\lambda_b\bar{\tau}_j) - \gamma(j+1,\lambda_b\bar{\tau}_{j-1})\Big], \qquad \text{(S27)}$$

and

$$\gamma(j,z) = (j-1)!\left[1 - e^{-z}\sum_{i<j}\frac{z^i}{i!}\right],$$

is the lower incomplete gamma function. For the $j = n_0$ term only, the upper argument $\lambda_b\bar{\tau}_{n_0}$ is to be replaced by zero, since that term is restricted to $o > \bar{\tau}_{n_0}$; the negative lower argument $\lambda_b\bar{\tau}_{n_0-1}$ is admissible as $\gamma(j,z)$ in the form above is entire.

As for the dark case, the mean detection time Eq. (S24) carries over under mismatch giving $\bar{t}_b = (\tau_{0,b}+\bar{o}')/(\lambda'_b\tau - 1)$ where $\bar{o}'$ follows from Eq. (S26) with $\lambda_b \to \lambda'_b$ and $\lambda_d \to \lambda'^{*}_b$. The dark rate enters $\bar{o}$ only as the conjugate of the bright rate so the true dark rate does not enter the bright-state statistics.

## III. ERROR RATE RELATIONS

### A. The two-boundary error relation

At the design rates the two achieved errors obey the exact relation

$$\frac{1-\epsilon_d}{\epsilon_d}\,\varepsilon_b + \varepsilon_d = 1, \qquad \text{(S28)}$$

which holds for the full two-boundary algorithm. It is not strictly needed but given here for completeness. When used in conjunction with Eq. (6) and $0 < \varepsilon_d < \epsilon_d/(1-\epsilon_d)$, it illustrates the second order accuracy of the single boundary results.

A complete record of the algorithm consists of the ordered photon arrival times $t_1 < \cdots < t_n$ and the termination time $T$, with $T$ and $n$ fixed by the stopping rule: $T = \tau_n$ for a dark declaration and $T = t_n < \bar{\tau}_n$ for a bright one. The admissible records form regions of the arrival-time space determined entirely by the thresholds. Under a Poisson process of rate $\lambda$ the probability density of a given record is

$$\lambda^n e^{-\lambda T}, \qquad \text{(S29)}$$

so the hypotheses enter only through this prefactor, while the regions and the integration variables $dt_1 \cdots dt_n$ are common to both. The ratio of the two densities on any record is

$$L_T = \frac{\lambda_d^n e^{-\lambda_d T}}{\lambda_b^n e^{-\lambda_b T}} = e^{-(\lambda_b-\lambda_d)(n\tau - T)}, \tag{S30}$$

the factor by which the record is more likely for a dark ion than for a bright one, where the conjugacy converts $(\lambda_d/\lambda_b)^n$ into $e^{-(\lambda_b-\lambda_d)n\tau}$. At a dark declaration $n\tau - T = n\tau - \tau_n = -\tau_{0,d}$ exactly, so

$$L_T = e^{(\lambda_b-\lambda_d)\tau_{0,d}} = \frac{1-\epsilon_d}{\epsilon_d}, \tag{S31}$$

a constant, because the dark boundary is met at an exact time.

The error relation now follows from normalization. Under either hypothesis the algorithm terminates with probability one. A dark ion meets a deadline even in the absence of the bright boundary, and the second boundary can only stop records sooner. Summing the dark-hypothesis density over all terminating records gives

$$1 = \sum_{\text{dark}} \lambda_d^n e^{-\lambda_d T} + \sum_{\text{bright}} \lambda_d^n e^{-\lambda_d T}, \tag{S32}$$

where the sums run over the dark- and bright-declaration records, including the integrals over their regions, and the second sum is by definition $\varepsilon_d$. In the first sum, Eq. (S30) exchanges the dark-hypothesis density for the bright-hypothesis density on each record, $\lambda_d^n e^{-\lambda_d T} = L_T\, \lambda_b^n e^{-\lambda_b T}$, and by Eq. (S31) the factor $L_T$ is the same constant on every dark record. Pulling it out, the remaining sum of the bright-hypothesis density over the dark records is $\varepsilon_b$, and Eq. (S28) follows with no approximation. The relation is the elementary instance, for exact boundary attainment, of the optional-stopping theorem for the likelihood-ratio martingale [5]; no property of $L_T$ beyond Eq. (S30) and Eq. (S31) has been used.

The derivation makes precise when Eq. (S28) survives rate mismatch. With true rates $(\mu, \mu^*)$ in place of $(\lambda_d, \lambda_b)$, the density ratio on a record is $(\mu/\mu^*)^n e^{(\mu^*-\mu)T}$, and the exchange step requires it to be constant on the dark records, where $n\tau - T = -\tau_{0,d}$; this holds precisely when $(\mu/\mu^*)^n = e^{-(\mu^*-\mu)n\tau}$, i.e. when the true rates are themselves a conjugate pair at the model spacing, in which case Eq. (S28) holds with $(1-\epsilon_d)/\epsilon_d$ replaced by $e^{(\mu^*-\mu)\tau_{0,d}}$. For generic mismatch the ratio acquires an $n$-dependence, no constant factor can be extracted, and each error channel must be evaluated from its own single-boundary closed form.

### B. Single-boundary identities

The same prefactor exchange applies within the single-boundary model, where it relates the closed-form distributions at conjugate rates. Both families have the form of a rate-independent volume multiplying Eq. (S29), so for any conjugate pair $\mu e^{-\mu\tau} = \mu^* e^{-\mu^*\tau}$ with $\mu\tau < 1 < \mu^*\tau$, the ratio of a distribution at the two rates is $e^{(\mu^*-\mu)(k\tau - t)}$ evaluated at the termination time. A dark declaration terminates exactly at $t = \tau_k$, where $k\tau - t = -\tau_{0,d}$; a bright declaration with $k$ photons terminates at $t < \bar{\tau}_k$, where $k\tau - t = \tau_{0,b} + (\bar{\tau}_k - t)$ exceeds $\tau_{0,b}$ by the overshoot. Hence

$$P_k\big|_{\mu^*} = e^{-(\mu^*-\mu)\tau_{0,d}}\, P_k\big|_{\mu}, \tag{S33}$$

and

$$e^{(\mu^*-\mu)(\bar{\tau}_k - t)}\, p_k(t)\big|_{\mu} = e^{-(\mu^*-\mu)\tau_{0,b}}\, p_k(t)\big|_{\mu^*}, \tag{S34}$$

are identical up to the overshoot factor, which is absent for a boundary met at an exact time and present for one crossed by a jump.

### C. Accuracy restrictions

One condition delimits the second-order accuracy of the single-boundary approximation for $\varepsilon_d$: the dark deadlines must lie beyond the bright deadlines requiring $\tau_{0,d} > \bar{\tau}_{n_0}$ or $\tau_{0,d} + \tau_{0,b} > n_0\tau$. If this is not the case, a single photon can simultaneously miss the dark deadline and meet a bright deadline, and the records retained by the single-boundary evaluation are overcounted at first order.

When $\epsilon_b \geq \epsilon_d$, this requires $\tau_{0,b}/\tau > n_0/2$. Since $n_0 = \lceil \tau_{0,b}/\tau \rceil$ implies $\tau_{0,b}/\tau > n_0 - 1 \geq n_0/2$ for $n_0 \geq 2$, the condition can fail only for $n_0 = 1$, where it reads

$$\frac{\lambda_b}{\lambda_d} < \frac{1-\epsilon_d}{\epsilon_d}\frac{1-\epsilon_b}{\epsilon_b} \tag{S35}$$

Violation therefore requires a contrast so high that a single photon flags the bright state and the second-order result is simply $\varepsilon_d \simeq 1 - e^{-\lambda_d \tau_{0,d}}$.

For $\epsilon_d > \epsilon_b$, violation can occur for $n_0 > 1$ at smaller contrasts and large asymmetry in the errors. Since a typical application would use either $\epsilon_d = \epsilon_b$ for simplicity, or $\epsilon_b > \epsilon_d$ to make the dark and bright errors equal, we do not consider $\epsilon_d > \epsilon_b$ regime further.

### D. Mismatch Approximations

The mismatched closed forms are functions of the relevant true rate and its conjugate, so their derivatives must account for the motion of the conjugate along the constraint $\mu e^{-\mu\tau} = \mu^* e^{-\mu^*\tau}$. Implicit differentiation, with the conjugacy used once more to trade the exponentials for the rates, gives

$$\frac{d\mu^*}{d\mu} = \frac{\mu^*(1-\mu\tau)}{\mu\,(1-\mu^*\tau)}, \quad \frac{d(\mu^*-\mu)}{d\mu} = \frac{\mu^*-\mu}{\mu\,(1-\mu^*\tau)}, \tag{S36}$$

negative for $\mu\tau < 1 < \mu^*\tau$: the two rates move toward each other along the conjugacy.

For the bright channel

$$\varepsilon_b' = e^{-(\lambda_b' - \lambda_b'^*)\tau_{0,d}} \tag{S37}$$

giving

$$\frac{d\ln\varepsilon_b'}{d\ln\lambda_b'} = -\frac{(\lambda_b' - \lambda_b'^*)\,\tau_{0,d}}{1-\lambda_b'^*\tau}, \tag{S38}$$

which, at the design point, is the exponent of the first of Eq. (10).

For the dark channel, we have

$$\varepsilon_d' = e^{-(\mu^*-\mu)(\tau_{0,d}+\bar{o}_{\rm v})}, \quad \bar{o}_{\rm v} = \frac{\ln\bar{\kappa}'}{\mu^*-\mu} \in (0,\tau) \tag{S39}$$

where $\mu = \lambda_d'$. Differentiating and writing the sensitivity in the same form as for the bright case gives

$$\frac{d\ln\varepsilon_d'}{d\ln\lambda_d'} = \frac{(\mu^*-\mu)(\tau_{0,b}+\bar{o}_{\rm s})}{\mu^*\tau - 1}, \tag{S40}$$

where

$$\bar{o}_{\rm s} = \bar{o}_{\rm v} - \mu\,(\mu^*\tau - 1)\,\frac{d\bar{o}_{\rm v}}{d\mu}, \tag{S41}$$

which, at the design point, is the exponent of the second of Eq. (10), where $\bar{o}_{\rm s}$ is the effective overshoot referred to there. The value overshoot, $\bar{o}_{\rm v}$, is confined to $(0,\tau)$ but $\bar{o}_{\rm s}$ is not: $\bar{o}_{\rm s}$ tracks $\bar{o}_{\rm v}$, and the bright-detection overshoot $\bar{o}$ of Eq. (S26), to within about $0.1\tau$ over a wide range of parameter space, departing where $\tau_{0,b}/\tau$ lies just below an integer at high contrast and the crossing migrates between the $n_0$ and $n_0+1$ photon channels. For direct evaluation, differentiating Eq. (8) gives the closed form

$$\frac{d\ln\varepsilon_d'}{d\ln\lambda_d'} = \frac{\lambda_d'\tau\,G + (1-\lambda_d'\tau)\,H}{1-(1-\lambda_d'\tau)\,G}, \tag{S42}$$

where

$$G = \sum_{m=0}^{n_0-1} \frac{(\lambda_d'\bar{\tau}_m)^m}{m!} e^{-\lambda_d'\bar{\tau}_m} \tag{S43}$$

and

$$H = \sum_{m=0}^{n_0-1} \frac{(\lambda_d'\bar{\tau}_m)^m}{m!} e^{-\lambda_d'\bar{\tau}_m}(\lambda_d'\bar{\tau}_m - m). \tag{S44}$$

## IV. AFTERPULSING: GENERAL CONSIDERATIONS

The afterpulse correction Eq. (13) applies at high contrast $x = \lambda_b/\lambda_d \gtrsim 100$ and $n_0 - \tau_{0,b}/\tau \gtrsim 1/\ln x$, which is relevant in many practical settings: applications aiming for high detection efficiency would typically have high contrast and one can always choose $\epsilon_b$ to keep clear of any step in $n_0$ with little compromise in performance. Moreover, the influence of afterpulsing diminishes with contrast. Nevertheless, a more general account is given for completeness.

As the contrast decreases or $\tau_{0,b}/\tau$ approaches an integer from below, the Erlang tail approximation breaks down and one must include times larger than $\bar{\tau}_{n_0}$. Since an afterpulse duplicates one count essentially instantly, a real photon stream $t_1 < t_2 < \cdots$ with a duplicate inserted at photon $j$ has every subsequent photon's count index raised by one. But $\bar{\tau}_{k+1} = k\tau - (\tau_{0,b} - \tau)$, which are the same deadlines with the offset shifted by $\tau$. So any affected record will record a bright state if and only if the real record triggers a bright state under the shifted deadlines. The afterpulsing correction is then

$$\Delta\varepsilon_d \approx p S_d^{\rm sh}\varepsilon_d^{\rm sh} \tag{S45}$$

with $\varepsilon_d^{\rm sh}$ and $S_d^{\rm sh}$ given by Eq. (8) and Eq. (S42), respectively, with both evaluated at $\lambda_d' = \lambda_d$ and offset $\tau_{0,b} - \tau$. Explicitly

$$S^{\rm sh} = \frac{\lambda_d\tau G^{\rm sh} + (1-\lambda_d\tau)H^{\rm sh}}{1-(1-\lambda_d\tau)G^{\rm sh}}, \tag{S46}$$

$$\varepsilon_d^{\rm sh} = 1-(1-\lambda_d\tau)G^{\rm sh} \tag{S47}$$

$$G^{\rm sh} = \sum_{m=0}^{n_0-2} \frac{(\lambda_d\bar{\tau}_{m+1})^m}{m!} e^{-\lambda_d\bar{\tau}_{m+1}}, \tag{S48}$$

$$H^{\rm sh} = \sum_{m=0}^{n_0-2} \frac{(\lambda_d\bar{\tau}_{m+1})^m}{m!} e^{-\lambda_d\bar{\tau}_{m+1}}\left(\lambda_d\bar{\tau}_{m+1} - m\right), \tag{S49}$$

with the upper limit $n_0-2$ taken to imply the result does not hold for $n_0 = 1$. In principle the $n_0 = 1$ result can be influenced by afterpulsing. A single photon that was too late to meet $\bar{\tau}_1$, can be early enough for the duplicate to beat $\bar{\tau}_2$. In this case

$$\Delta\varepsilon_d = p(e^{-\lambda_d\bar{\tau}_1} - e^{-\lambda_d\bar{\tau}_2}). \tag{S50}$$

To understand Eq. (S45), first note that a rate $\lambda$ only enters the probability density for any record through the factor $\lambda^n e^{-\lambda T}$, for which

$$\lambda\frac{d}{d\lambda}\left(\lambda^n e^{-\lambda T}\right) = (n-\lambda T)\left(\lambda^n e^{-\lambda T}\right). \tag{S51}$$

It then follows that

$$S_d = \frac{d\ln\varepsilon_d}{d\ln\lambda_d} = \langle n - \lambda_d T\rangle_{\rm err}, \tag{S52}$$

where the average is over the erring-records, which are dark records terminating with a bright declaration. Every erring record satisfies $n\tau = T + \tau_0 + o$ with $o \in (0,\tau)$, so averaging over the ensemble together with Eq. (S52) to eliminate $\langle T\rangle_{\rm err}$ yields

$$S_d \;<\; \langle n\rangle_{\rm err} \;<\; \frac{S_d}{1-\lambda_d\tau}, \tag{S53}$$

which holds at any offset. At any appreciable contrast $S_d$ is then an estimate for $\langle n\rangle_{\rm err}$ and Eq. (S45) the natural extension to Eq. (13).

The bounds in Eq. (S53) can be closed. The overshoot density of the erring ensemble follows from the diagonal sum of Sec. II C evaluated at the dark rate: with $x = \lambda_d e^{-\lambda_d \tau}$ the principal root is $\zeta = \lambda_d$ itself, the exponent $(\lambda_d - \zeta)\, o$ of the main term vanishes, and, on $(0, \tau)$,

$$f_{\rm err}(o) = A_d - e^{\lambda_d(\tau_0+o)} {\sum_{j=1}^{n_0}}' \left(\lambda_d e^{-\lambda_d \tau}\right)^j \frac{(\bar{\tau}_j - o)^{j-1}}{(j-1)!}, \qquad \text{(S54)}$$

with

$$A_d = \lambda_d\, e^{\lambda_d \tau_0}\, Q\big(\lambda_d e^{-\lambda_d \tau}\big) = \frac{\lambda_d\,(1-\varepsilon_d)}{1-\lambda_d \tau}, \qquad \text{(S55)}$$

where the prime carries the same restriction as in Eq. (S25) and the second equality in Eq. (S55) follows from Eq. (8). The density is unnormalized, $\int_0^\tau f_{\rm err}\, do = \varepsilon_d$, which recovers Eq. (8). Term-by-term integration, exactly as for Eq. (S26), gives

$$\varepsilon_d \,\langle o \rangle_{\rm err} = A_d\, \frac{\tau^2}{2} - \sum_{j=1}^{n_0} \bar{o}_j \Big|_{\lambda_b \to \lambda_d}, \qquad \text{(S56)}$$

with $\bar{o}_j$ as in Eq. (S27) and the $j = n_0$ argument replaced by zero as there. Averaging the crossing constraint and eliminating $\langle T \rangle_{\rm err}$ then determines the multiplicity exactly,

$$\langle n \rangle_{\rm err} = \frac{S - \lambda_d\big(\tau_0 + \langle o \rangle_{\rm err}\big)}{1 - \lambda_d \tau}, \qquad \text{(S57)}$$

$$\langle T \rangle_{\rm err} = \langle n \rangle_{\rm err}\, \tau - \tau_0 - \langle o \rangle_{\rm err}, \qquad \text{(S58)}$$

which satisfies Eq. (S53) identically.

## V. DECAY COEFFICIENT

For the decay corrections Eq. (12), a decay at time $s$ flips the eventual declaration unless the posterior is already near the original boundary. Here $T$ denotes the time the record would have terminated absent decay, so $L_T$ is the fixed boundary value of Eq. (S31) rather than a second running variable. The exception is weighted by the likelihood-ratio value at the switch: for a dark ion, $\mathbb{E}_d[L(s)]$ grows at the rate $c = (\lambda_b - \lambda_d)^2/\lambda_b$, so the boundary factor $\epsilon_d/(1-\epsilon_d)$ cancels against $L_T = (1-\epsilon_d)/\epsilon_d$ (Eq. (S31)), and the deficit integrates to the fixed time $1/c = \lambda_b/(\lambda_b - \lambda_d)^2$. The bright channel follows by the same argument with $\Lambda = 1/L$, whose boundary factor $\epsilon_b/(1-\epsilon_b)$ cancels against $\Lambda_T = (1-\epsilon_b)/\epsilon_b$, and growth rate $(\lambda_b - \lambda_d)^2/\lambda_d$ under the bright hypothesis.

## VI. REFERENCE IMPLEMENTATIONS FOR SEQUENTIAL DEADLINE DETECTION

The decision rule requires only a photon counter (here the rising-edge output of a single-photon counting module, SPCM), a timer, and the three precomputed integer constants $\tau$, $\tau_{0,d}$, and $\tau_{0,b}$ (in clock cycles). No floating-point arithmetic or multiplication is required: each running deadline is maintained by a single addition of $\tau$ per photon arrival. A basic pseudocode description of the algorithm is given in Algorithm 1. Our implementation uses a Xilinx Artix-7 FPGA clocked at 100 MHz, but the logic is sufficiently simple that a much smaller FPGA, or even a CPLD, would suffice.

Algorithm 1. Sequential deadline detection gateware implementation

```
pre-defined constants: τ, τ_{0,d}, τ_{0,b}
t ← 0                                              ▷ elapsed time
n ← 0                                              ▷ photon count
D ← τ_{0,d}                                        ▷ dark deadline
B ← −τ_{0,b}                                       ▷ bright deadline
loop                               ▷ one iteration per clock cycle
    if rising edge on SPCM input this cycle then
        n ← n + 1
        B ← B + τ
        if t < B then
            return (BRIGHT, n, t)
        end if
        D ← D + τ
    else if t ≥ D then
        return (DARK, n, t)
    end if
    t ← t + 1
end loop
```

We additionally provide an implementation in ARTIQ (Advanced Real-Time Infrastructure for Quantum physics) [6], an open-source platform widely used in the atomic physics community, for the convenience of readers who use this platform and would like to implement the algorithm without low-level FPGA programming. A basic kernel is given in Listing 1. It can be readily adapted to include a deadtime that rejects counts arriving within a fixed interval of the previous count, mitigating afterpulsing errors as described in the main text.

We note that the ARTIQ implementation incurs an additional latency after the detection decision has been reached, primarily from the `break_realtime` call required to resynchronise the timeline cursor with the wall clock. This overhead is negligible in most settings, but where the detection duty cycle is the limiting factor the low-level implementation of Algorithm 1 should be preferred.

Listing 1. ARTIQ kernel applying the same decision rule as Algorithm 1, tested with ARTIQ 7 and later.The kernel is event-driven: `timestamp_mu` returns the timestamp of the next photon, or −1 if the running dark deadline is reached first.

```
import numpy as np
from artiq.experiment import *

@kernel
def detect(self, ttl_in, tau, tau_0d, tau_0b):
    """
    ttl_in is the TTL channel (ARTIQ core driver TTLInOut) connected to SPCM
    tau, tau_0d, tau_0b are precomputed durations in machine units (int64)
    """
    BRIGHT = 1
    DARK = 0

    t_start = now_mu()

    # Initialize variables and deadlines
    n = np.int32(0)
    state = DARK
    last_ts = np.int64(0)
    dark_mu = t_start + tau_0d
    bright_mu = t_start - tau_0b

    # Open gate to start logging photon arrivals
    ttl_in._set_sensitivity(1)

    while True:
        # Return the next photon's timestamp, or -1 if the timer reaches dark_mu
        ts = ttl_in.timestamp_mu(dark_mu)
        if ts < 0:
            last_ts = dark_mu
            state = DARK
            break

        # Increment photon number and deadlines
        n += 1
        dark_mu += tau
        bright_mu += tau

        if ts < bright_mu:
            last_ts = ts
            state = BRIGHT
            break

    # Close gate and break realtime to advance time cursor
    self.core.break_realtime()
    ttl_in._set_sensitivity(0)
    ttl_in.count(now_mu())  # empty FIFO buffer

    return (state, n, last_ts - t_start)
```



\end{document}